\documentclass[10pt,letterpaper]{article}
\usepackage[top=0.85in,left=2.75in,footskip=0.75in,marginparwidth=2in]{geometry}

\usepackage[utf8]{inputenc}

\usepackage{cite}

\usepackage{nameref,hyperref}

\usepackage[right]{lineno}

\usepackage{microtype}
\DisableLigatures[f]{encoding = *, family = * }

\raggedright
\usepackage{changepage}

\usepackage[aboveskip=1pt,labelfont=bf,labelsep=period,singlelinecheck=off]{caption}

\makeatletter
\renewcommand{\@biblabel}[1]{\quad#1.}
\makeatother

\usepackage{lastpage,fancyhdr,graphicx}
\usepackage{epstopdf}
\fancyheadoffset[L]{2.25in}
\fancyfootoffset[L]{2.25in}

\usepackage{color}

\definecolor{Gray}{gray}{.25}

\usepackage{graphicx}

\usepackage{sidecap}

\usepackage{wrapfig}
\usepackage[pscoord]{eso-pic}
\usepackage[fulladjust]{marginnote}
\usepackage[table,xcdraw]{xcolor}
\reversemarginpar

\begin{document}
\vspace*{0.35in}

\begin{flushleft}
{\Large
\textbf\newline{Minimizing cyber sickness in head mounted display systems: design guidelines and applications}
}
\newline
\\
Thiago M. Porcino\textsuperscript{1},
Esteban W. Clua\textsuperscript{1},
Cristina N. Vasconcelos\textsuperscript{1},
Daniela Trevisan\textsuperscript{1},
Luis Valente\textsuperscript{1}

\bigskip
\bf{1} Federal Fluminense University, Brazil
\\
\bigskip
* thiagomp@ic.uff.br

\end{flushleft}

\section*{Abstract}
We are experiencing an upcoming trend of using head mounted display systems in games and serious games, which is likely to become an established practice in the near future. While these systems provide highly immersive experiences, many users have been reporting discomfort symptoms, such as nausea, sickness, and headaches, among others. When using VR for health applications, this is more critical, since the discomfort may interfere a lot in treatments. In this work we discuss possible causes of these issues, and present possible solutions as design guidelines that may mitigate them. In this context, we go deeper within a dynamic focus solution to reduce discomfort in immersive virtual environments, when using first-person navigation. This solution applies an heuristic model of visual attention that works in real time. This work also discusses a case study (as a first-person spatial shooter demo) that applies this solution and the proposed design guidelines.


\section{Introduction}

Head Mounted Displays (HMDs) are proving to be an important tool for an increase in 3D games immersion. Currently, a large and rapid growth in the emergence of several solutions can be observed and is being pointed as one of the most important game industry trends for the next years. Although the entertainment industry is being strongly influenced by the field, serious applications are becoming one of the most important topics for this context, proving to be an important tool for health care, treatments and scientific visualization. 

Latency was a serious problem in previous generation of HMDs, making it difficult to produce games for the mass market ~\cite{oculusblog}. Modern HMDs assemblers state that they bypassed this issue, pointing that VR could now become the next generation gaming platform.

However, many users have reported discomfort due to the prolonged use of VR devices with other different causes, rather than latency ~\cite{carnegie2015reducing}. According to ~\cite{duchowski2014reducing} , ~\cite{shibata2011zone} and ~\cite{yano2004two}, the causes for visual discomfort regarding stereo vision devices can be listed as:

\begin{itemize}
 \item Eyewear with image separation between eyes; 
 \item Incorrect calibration or poor focus simulation;
  \item Convergence accommodation conflict. 
\end{itemize}

While the discomfort is an important issue for entertainment purposes, in health games and applications it becomes critical, since patients may already be in uncomfortable or unnatural situations. This paper addresses the discomfort caused by many factors in stereoscopic Head Mounted Displays with head movement tracking. Different factors can contribute for health problems in HMDs. While in humans the focus produces blurriness effects according to the objects in the visual area ~\cite{zhang2014human}, researches try to simulate the focus selection in HMDs. Popular HMDs doesn't have an eye-tracking system, in order that, they don't produce an adequate simulation of depth of field and focus selection.

Moderns Head Mounted Displays resolves many issues of earlier devices. But, even with a lack of good hardware implementation, improperly design content will contribute to an unhealthy experiences. 

Different depth of field (DoF) simulation techniques were proposed in computer graphics, in order to generate realistic scenes ~\cite{kottravel2015coverage}. Virtual reality and general entertainment applications often use such techniques to grab the user’s attention and enhance immersion ~\cite{li2011study}.

In this work we propose a design guideline to help developers to minimize sickness problems in HMD systems. Furthermore, we develop a game prototype that will publish as case of study in public domain and tries to make a deeper analysis into the focus selection features.

\section{Related Work}

Although the importance of the topic, the literature still lack of a complete bibliography for this subject, being most of the works empirical observations and reports. 

Concerning cybersickness research, Davis et al. ~\cite{davis2015comparing} designed tests to evaluate two different versions of the Oculus Rift device (i.e. the DK1 and DK2 models). These tests contained tasks that induced users to feel discomfort while using these devices. The evaluation had 30 users experience two different scenarios, both based on a roller-coaster simulation. Davis et al.~\cite{davis2015comparing} concluded that more complex and more realistic scenarios caused higher levels of discomfort.

Tan et al.~\cite{tan2015exploring} focused on game experience using HMDs, evaluating 10 participants in a  first-person shooter game using an Oculus Rift DK1 device. The evaluation compared game playing with the HMD and without it. Tan et al.~\cite{tan2015exploring} reported that most participants felt higher levels of discomfort while playing with the HMD.

The work by Carnegie and Rhee \cite{carnegie2015reducing} is closely related to our work, as they also aimed at reducing discomfort in HMD systems. More specifically, they proposed a GPU based DoF simulation to decrease discomfort in HMD systems. Instead of using eye tracking systems (a common solution to estimate focus areas precisely), they developed a dynamic real time DoF solution to keep the screen center in focus. Their system used HMD devices that react to neck movements. As a result, the user is able to change focus when moving his/her head. However, to mitigate discomfort that arises in sudden focus change, the system introduces a 500ms delay before updating the images in the HMD device. According to Carnegie and Rhee \cite{carnegie2015reducing}, the time required to reorient focus depends on the user’s age and the lighting conditions in the scene. They assume that for a real time performance, users require 500 ms to refocus from an infinite distance to approximately 1 meter. To assess their solution, they evaluated 20 participants with a sickness simulator questionnaire. For each of its 18 questions, participants verbally responded to symptoms using a Likert 5-point scale (ranging from ”none” -- without symptoms --  to ”severe”, meaning traumatic symptoms). This assessment revealed that 30\% of users who used a HMD system for the first-time were unable to use the system for more than 30 minutes, due to discomfort issues. When compared to these findings, our results demonstrated that DoF rendering techniques can significantly reduce the discomfort that users experience in applications with HMDs. Although we initially used the adaptation time that Carnegie and Rhee ~\cite{carnegie2015reducing} suggested (i.e. 500ms), in our tests we used empirical observations to determine the the focus adaptation time.

In an earlier work ~\cite{porcino2016dynamic}, we developed a model for dynamic selection for focusing virtual objects with a dynamic ROI (region of interest) in real time. By dynamic, we mean that our model moves the ROI in the 3D scene during the application lifetime. This model simulates a self-extracting mechanism of visual focus, which isolates ROIs in the visual field. Our model uses the ROIs to calculate DoF effects in real time, decreasing discomfort in HMD usage. 

Cybersickness in applications that use HMDs is still an open problem according to the literature ~\cite{carnegie2015reducing}. Generally, the available research works aim at analysing the reasons that generate discomfort symptoms. In this regard, we used the insights we had in our earlier work~\cite{porcino2016dynamic} and the experiences related in the literature~\cite{balk2013simulator} to design a guideline that may minimize cybersickness in HMD applications. This is not a complete guideline and there is still many other aspects that must be included in near future.

\section{Our Guideline}

There are several factors that contribute to causing sickness and discomfort in head mounted display systems ~\cite{carnegie2015reducing},~\cite{yaooculus}. This discomfort may lead to stronger effects, such as nausea, eyestrain, headache, and vertigo~\cite{kennedy1993simulator}. Academic literature on guidelines to avoid these symptoms are scarce. Currently, the most comprehensive list of guidelines are takes form the “Oculus Best Practices” ~\cite{yaooculus}, which is an information source compiled after empirical practice with the Oculus Rift device. Other HMD assemblers are also compiling their own documents.

According to the current literature ~\cite{so2001effects}, ~\cite{lin2004virtual} ,~\cite{draper2001effects}, ~\cite{kolasinski1995simulator} numerous factors can contribute to generate sickness in these environments and can be solved through software development, such as:

\begin{itemize}

\item Acceleration - High acceleration movements in VR is a strong factor that contributes to discomfort. On the other hand, lower acceleration movements generate  more comfortable experiences~\cite{so2001effects}. This occurs because increasing acceleration can generate sensory conflicts that cause disparities in the brain. According to ~\cite{technologyreview}: “The issue is constant deceleration and acceleration. It’s actually the duration of that change, rather than the magnitude, that makes people change. An instant acceleration from zero to 100, like truly instant, actually makes very few people ill. But slowly ramping it up and then ramping it down is a lot more uncomfortable for a lot of people."

\item Degree of control - cameras movements not initiated by the user can contribute to generate sickness. In other words, unexpected camera action (movement or cessation of movement) can be uncomfortable for the user. Anticipating and preparing the user’s visual motion experience.  According to Lin et al.~\cite{lin2004virtual}: “Having an avatar that foreshadows impending camera movement can help users anticipate and prepare for the visual motion, potentially improving the comfort of the experience”.

\item Duration use time - Discomfort levels increase proportionally to device usage time. Hence, the application should allow users to pause the experience for a rest, and then resume it later. Conversely, an application could suggest users to take a break periodically, to avoid discomfort symptoms.

\item Field of View - According to ~\cite{yaooculus} there are two kinds of field of view: The Display FOV is  the area of the visual field composed by the display and the Camera FOV is the virtual environment area where engine draws to the display (cFOV). Reducing display FOV can reduce simulator sickness~\cite{draper2001effects}, but also reduces the level of immersion. The manipulation of camera FOV, which causes an unnatural movement of the user inside the virtual environment, can produce discomfort. 

\item Jumping movements - Realistic body movements, when a first person style game is being used, is common in games and simulators. However, since in VR the movements of the virtual avatar is not necessarily being made by the real user, the lack of this movements can bring serious disparity. In order to avoid this issue, it may be important to include movements based on jumps instead of continuous walks. 

\item Latency -  Latency was a serious problem in previous generation of HMDs ~\cite{olano1995combatting}, making it difficult to produce games for the mass market~\cite{oculusblog}. It is crucial to ensure that the VR application does not drop frames or produces delays. In many games, the frame rate slows down when these applications process complex graphics elements. While this is not a severe issue in non-VR games, in VR systems with HMD displays this issue contributes to creating discomfort in users ~\cite{draper2001effects} ~\cite{kolasinski1995simulator}.

\end{itemize}

In computer generated graphics, we can say that an application usually displays the virtual scene with “infinity focus”, which means that the system forms images of objects at an infinite distance away. For the HMD end-user, this means that all objects appears to be focused, which is a state that contributes to create discomfort. In this work, we propose a solution to mitigate this situation through five design guidelines. When applying these guidelines, an application is able to implement a dynamic focus model. Our proposal guidelines for this are:

\begin{itemize}

\item Blur level: The blur level in the scene needs to be subtle enough to match the depth effect that occurs in real life. If the blur level is too high, users are likely to experience discomfort; 

\item Refocus time: A recent study~\cite{carnegie2015reducing} has demonstrated that the time that that an individual takes to change focus from different objects (i;e. the transition focus time) varies for each person. For real time applications, we assume that users will feel comfortable if the transition focus speed is at least 500 milliseconds ~\cite{carnegie2015reducing}.
        
\item Observer’s attention: Although most people tend to focus at the center of the screen~\cite{carnegie2015reducing}, people can place focus anywhere, according to the object of interest. This means that focused objects will not necessarily be at the center of the screen.     

\item Depth: People are able to focus objects in different depth levels (i.e. place focus selectively on near or far objects at will). So, a virtual world should also simulate this effect by enabling users to alternate focusing of near and far objects.

\item Persistence: In real world, humans can maintain perception in moving objects. To simulate this effect in virtual environments with HMDs,  the system should enable the user to maintain focus even further after an object reaches the focal point.  
\end{itemize}

\section{Dynamic Focus Selection Model}
\label{sec:3}

\begin{figure}[htb]
	\begin{center}
			\includegraphics[width=2.76in]{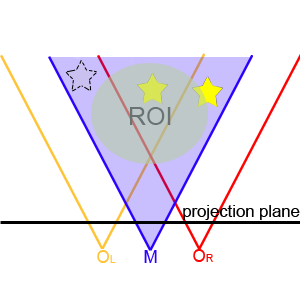}
	\end{center}
	\caption{\textbf{OL}, \textbf{OR}, and their corresponding middle point \textbf{M}. ROI associated with the visual field pre-selects objects considered as candidates of visual attention (filled objects)}
	\label{fig:1}
\end{figure}

Among all the guidelines proposed in the previous session, we give special attention to the focus selection process. We previously proposed a model that uses a pair of stereo cameras (\textbf{CL} and \textbf{CR}), whose optical centers (\textbf{OL} and \textbf{OR}) were positioned as left and right, to align both camera image planes and epipolar, producing IL and IR images ~\cite{porcino2016dynamic}. Figure ~\ref{fig:1} illustrates this model, where a virtual camera (\textbf{CM}) is placed at the midpoint (\textbf{M}) of the line connecting the two projection centers (\textbf{OL}) and (\textbf{OR}). The normal vectors are parallel to both projection planes. \textbf{M} is associated with the normalized vector (\textbf{N}) defined by direction such that \textbf{CM} aligns with the parallel geometry. 

The model suggest the object with best focus in the scene by using an heuristic with weights and a region of interest. Using a region of interest (ROI) allows optimizing real time calculations in complex scenes, by excluding objects located outside the user visual field.

After this first step, all objects that remain completely or partially in the ROI become possible targets of the users’ visual focus and thus are to be viewed in focus. An importance metric function \textbf{I(o, CM)} (responsible for choose the focus object) analyses and select objects to receive focus , where \textbf{"o"} represents a given object within the ROI and \textbf{CM} is a set of cameras. 

\subsection{Heuristics For First-Person Virtual Environment Navigation}

The heuristic model has an importance function (responsible for choose the focus object) for first-person exploration of virtual environments using HMDs (Equation ~\ref{eq1}).

\begin{equation}
\label{eq1}
I(o,C_{1}, C_{2}) =  (PRM * RM(o, C_{M})) + (PD * D(o, C_{M})) + (PV * V(o))      
\end{equation} 

\noindent where \textbf{PRM + PD +  PV} = 1 represents weight factors among the metrics  \textbf{RM}, \textbf{D}, and \textbf{V}.

\begin{figure}[htb]
	\begin{center}
			\includegraphics[width=2.76in]{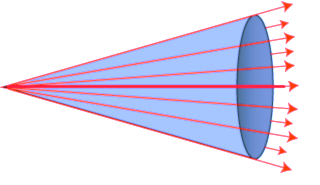}
	\end{center}
	\caption{Metric Rays.}
	\label{fig:2}
\end{figure}

The first visual focus evaluation parameter considers objects closer to the camera center as more important. The first evaluation parameter (metric rays) is achieved by a series of rays within a cone centered at the midpoint between \textbf{C1} and \textbf{C2} (illustrated in figure ~\ref{fig:2}). The  interior cone is divided into concentric layers \textbf{k}. For each layer, \textbf{n} sets of rays are generated, called metric rays \textbf{RM}. From their common origin \textbf{M} , their uniform dispersion is then defined by the golden rule, resulting in \textbf{k} multiplied per \textbf{n} rays.

The heuristic model evaluates the number of metric rays hitting a scene object. Each ray has an alpha weight depending on how close each layer is relative to the cone’s center, considering the importance of objects centered in the scene in the heuristic calculation (Equation ~\ref{eq2}).

\begin{equation}
\label{eq2}
RM(CM,o)=\sum_{i=1}^{k}\sum_{j=1}^{n}  c(j,o)* \alpha(i)
\end{equation}

The binary function \textbf{c} determines whether rays \textbf{j} collide with objects \textbf{o}. If a given radius encompasses more than one object in the scene, the heuristic considers only those that are closest to the camera. The second parameter \textbf{D(CM,o)}  uses the depth relative to point \textbf{M} of a specific object \textbf{o} belonging to a ROI. This metric evaluates the distance between an object and the user, assuming that closest elements tend to receive more attention, so become focused.

The third parameter is the added value of the object \textbf{V} and is incorporated to contextualize specific applications. For example, in games, dangerous or beneficial objects deserve greater attention from users compared with those that are merely decorative. Some objects can therefore be unconventionally focused. “Added value” is thus individually factored for each scene object by application models.

In a game, for instance, a dangerous or beneficial object deserves more observer attention than those that are purely decorative. Thus, depending on the context, objects can be focused in a non-conventional manner. The “earned value” technique should be individually attributed to each object on a scene by either an application modeler.

In order to use our heuristic for the suggested guideline, we propose that the designers of the application create a set of importance rates for the objects. The higher the rate, the stronger will be chance that the element is selected for focus. In this sense, the designer should categorize and create groups of candidates for focus.


\section{Case of Study: Dynamic Focus Selection}
\label{sec:4}

We developed  a virtual space shooter game demo to validate our design guidelines, using the Unity 3D game engine and the Oculus Rift HMD. The user is able to activate or deactivate the dynamic focus plugin during gameplay, through a keyboard key. Researchers can use this demo to detect sickness issues when users wear HMDs for a established time per session. As far as we know, this is the first playable game which uses an dynamic focus selection connected to the user’s attention without any system of eye tracking.

\begin{figure}[htb]
	\begin{center}
			\includegraphics[width=2.76in]{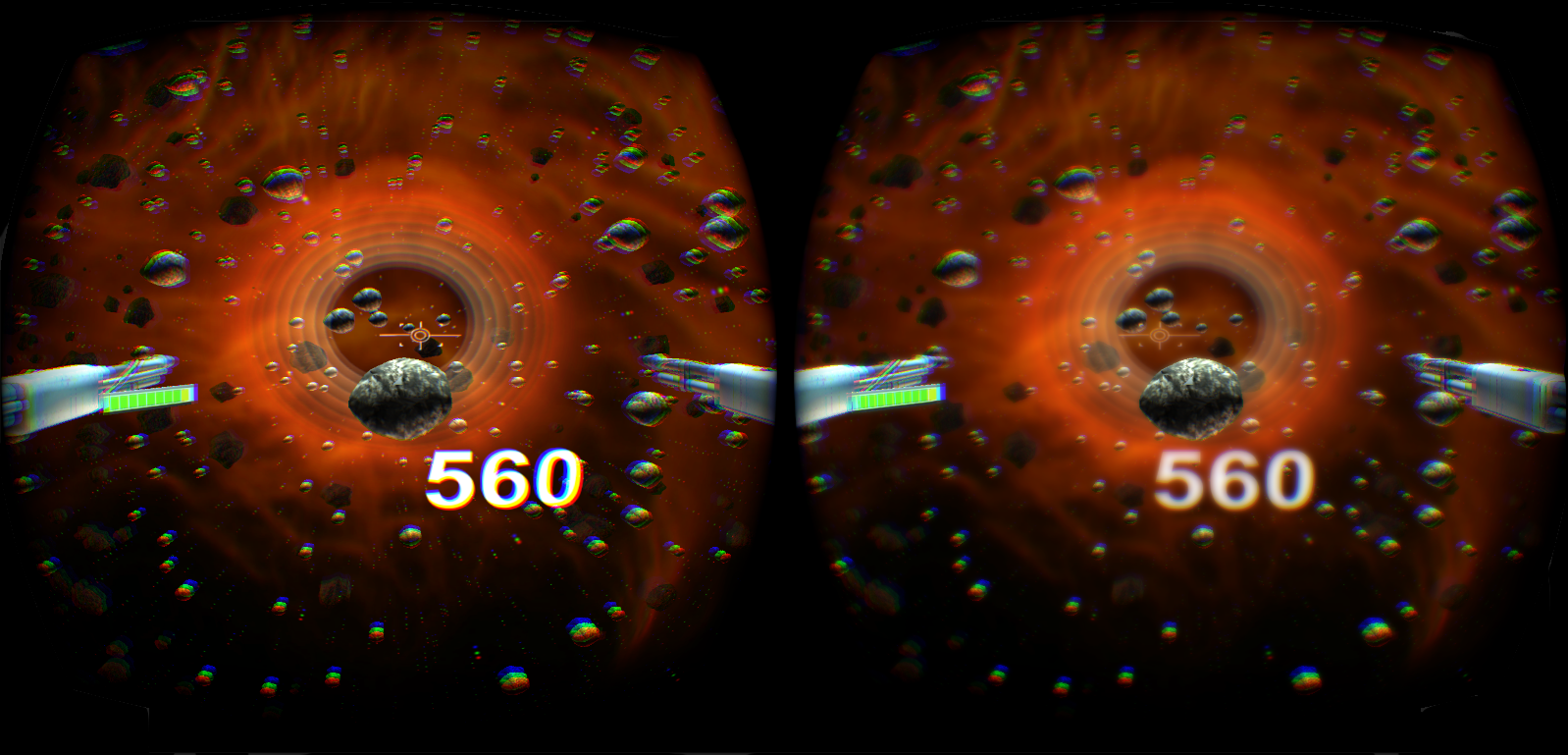}
	\end{center}
	\caption{Virtual space shooter game demo. The dynamic focus selection was enabled only in the right side of the image.}
	\label{fig:3}
\end{figure}

The virtual environment contains an interactive scenario (Figure \ref{fig:3}). The user controls a ship and must destroy as many asteroids as possible. The user earns 20 points for each asteroid he destroys. If an asteroid hits the user ship, the user receives one damage point. If the user receives 10 damage points, the game is over and the application informs the user his maximum score. In this demo, the user progresses to different game levels according to his score, as summarized by Table ~\ref{table1}.

\begin{table}[ht]
\centering
\caption{Levels score conditions}
\label{table1}
\begin{tabular}{|l|l|}
\hline
\multicolumn{1}{|c|}{\textbf{Level}} & \multicolumn{1}{c|}{\textbf{Score}} \\ \hline
1                                    & less than 500 points                \\ \hline
2                                    & between 500 and 1000 points         \\ \hline
3                                    & between 1001 and 2000 points        \\ \hline
4                                    & between 2001 and 3000 points        \\ \hline
5                                    & between 3001 and 5000 points        \\ \hline
6                                    & more than 5000 points               \\ \hline
\end{tabular}
\end{table}

\section{Test Protocol}

We designed a test protocol to help other researchers to reproduce our usability tests. Firstly, we recommend using these two questionnaires:

\begin{itemize}
\item SSQ -  A simulator sickness questionnaire. A recent example of SSQ can be found in ~\cite{balk2013simulator}, while an earlier example can be found in~\cite{kennedy1993simulator}. 

\item Profile Questionnaire - A questionnaire containing  name, age, gender and academic background of each candidate.
\end{itemize}

The SSQ is widely used to describe and assess simulator sickness. The questionnaire asks users to score 16 symptoms on a four point scale from zero to three~\cite{balk2013simulator}. 

Recent SSQ usability tests used 3 SSQ~\cite{porcino2016dynamic} in the following order:
\begin{itemize}

\item SSQ applied before exposition.

\item SSQ applied after first session.

\item SSQ applied after second session.

\end{itemize}

After that, the researcher should collect all questionnaires (1 profile questionnaire and 3 SSQ from each user).

The evaluation test can consists of the six sequentially tasks, as proposed in~\cite{porcino2016dynamic}:

\begin{itemize}
\item Filling a profile questionnaire;
\item Filling an SSQ;
\item Completion of the first test session using an HMD device;
\item Filling an SSQ; 
\item Completion of the second test session using an HMD device;
\item Filling an SSQ. 
\end{itemize}

After the evaluation tests the researcher calculates the scores to analyse the discomfort issues.

\section{Data Analysis (SSQ Calculation of Scores)}

Balk ~\cite{balk2013simulator} classifies the 16 symptoms evaluated in the SSQ into three general classes: Nausea, Oculomotor and Disorientation. Weights are attributed to each of the classes and summed along to acquire a unique score. We design a compact table that considers all parts of the evaluation tasks. Table ~\ref{table2}~\cite{porcino2016dynamic} contains  three questionnaire columns (Q1, Q2 and Q3), which use a four point scale (ranging from zero to three).

Table ~\ref{table3} shows the weight distribution considering these three classes ~\cite{balk2013simulator}. The point scales of symptoms belonging to each class (nausea, oculomotor and disorientation) are summed along each column. After that, the results of each column (nausea, oculomotor and disorientation) are multiplied to an individual value, where 9.54 refers to Nausea, 7.58 to oculomotor and 13.92 to Disorientation. Besides that, there is a total score which is the sum of all point scales of symptoms belonging to the three symtoms classes (nausea, oculomotor and disorientation) and multiplied for the value of 3.74~\cite{kennedy1993simulator}.

\begin{table}[ht]
\centering
\caption{SSQ evaluation}
\label{table2}
\begin{tabular}{|l|
>{\columncolor[HTML]{EFEFEF}}l |
>{\columncolor[HTML]{EFEFEF}}l |
>{\columncolor[HTML]{EFEFEF}}l |
>{\columncolor[HTML]{EFEFEF}}l |
>{\columncolor[HTML]{C0C0C0}}c |
>{\columncolor[HTML]{C0C0C0}}l |
>{\columncolor[HTML]{C0C0C0}}l |
>{\columncolor[HTML]{C0C0C0}}l |
>{\columncolor[HTML]{9B9B9B}}c |
>{\columncolor[HTML]{9B9B9B}}l |
>{\columncolor[HTML]{9B9B9B}}l |
>{\columncolor[HTML]{9B9B9B}}l |}
\hline
\textbf{}                                                       & \multicolumn{4}{c|}{\cellcolor[HTML]{EFEFEF}{\color[HTML]{000000} \textbf{Q1}}} & \multicolumn{4}{c|}{\cellcolor[HTML]{C0C0C0}{\color[HTML]{000000} \textbf{Q2}}}                                                                      & \multicolumn{4}{c|}{\cellcolor[HTML]{9B9B9B}{\color[HTML]{000000} \textbf{Q3}}}                                                                      \\ \hline
\textit{Symptoms}                                                      & 0                  & 1                  & 2                 & 3                 & \multicolumn{1}{l|}{\cellcolor[HTML]{C0C0C0}0} & 1                                              & 2                                              & 3 & \multicolumn{1}{l|}{\cellcolor[HTML]{9B9B9B}0} & 1                                              & 2                                              & 3 \\ \hline
\begin{tabular}[c]{@{}l@{}}1. General \\ discomfort\end{tabular}       &-                  &-                  &-                 &-                 &-                                              &-                                              &-                                              &- &-                                              & \multicolumn{1}{c|}{\cellcolor[HTML]{9B9B9B}-} &-                                              &- \\ \hline
2. Fatigue                                                             &-                  &-                  &-                 &-                 &-                                              &-                                              &-                                              &- &-                                              & \multicolumn{1}{c|}{\cellcolor[HTML]{9B9B9B}-} &-                                              &- \\ \hline
3. Headache                                                            &-                  &-                  &-                 &-                 &-                                              & \multicolumn{1}{c|}{\cellcolor[HTML]{C0C0C0}-} &-                                              &- &-                                              & \multicolumn{1}{c|}{\cellcolor[HTML]{9B9B9B}-} &-                                              &- \\ \hline
4. Eye strain                                                          &-                  &-                  &-                 &-                 &-                                              &-                                              &-                                              &- &-                                              &-                                              &-                                              &- \\ \hline
\begin{tabular}[c]{@{}l@{}}5. Difficult \\ focusing\end{tabular}       &-                  &-                  &-                 &-                 &-                                              & \multicolumn{1}{c|}{\cellcolor[HTML]{C0C0C0}-} &-                                              & - & -                                              & \multicolumn{1}{c|}{\cellcolor[HTML]{9B9B9B}-} &-                                              &- \\ \hline
\begin{tabular}[c]{@{}l@{}}6. Increased \\ salivation\end{tabular}     &-                  &-                  &-                 &-                 &-                                              &-                                              &-                                              &- &-                                              &-                                              &-                                              &- \\ \hline
7. Sweating                                                            &-                  &-                  &-                 &-                 &-                                              & \multicolumn{1}{c|}{\cellcolor[HTML]{C0C0C0}-} &-                                              &- &-                                              &-                                              & \multicolumn{1}{c|}{\cellcolor[HTML]{9B9B9B}-} &- \\ \hline
8. Nausea                                                              &-                  &-                  &-                 &-                 &-                                              & \multicolumn{1}{c|}{\cellcolor[HTML]{C0C0C0}-} & \multicolumn{1}{c|}{\cellcolor[HTML]{C0C0C0}-} &- &-                                              & \multicolumn{1}{c|}{\cellcolor[HTML]{9B9B9B}-} &-                                              &- \\ \hline
\begin{tabular}[c]{@{}l@{}}9. Difficulty \\ concentrating\end{tabular} &-                  &-                  &-                 &-                 &-                                              & \multicolumn{1}{c|}{\cellcolor[HTML]{C0C0C0}-} &-                                              &- &-                                              & \multicolumn{1}{c|}{\cellcolor[HTML]{9B9B9B}-} & \multicolumn{1}{c|}{\cellcolor[HTML]{9B9B9B}-} &- \\ \hline
\begin{tabular}[c]{@{}l@{}}10. "Fullness \\ of head"\end{tabular}      &-                  &-                  &-                 &-                 & \multicolumn{1}{l|}{\cellcolor[HTML]{C0C0C0}-} &-                                              &-                                              &- & \multicolumn{1}{l|}{\cellcolor[HTML]{9B9B9B}-} &-                                              &-                                              &- \\ \hline
\begin{tabular}[c]{@{}l@{}}11. Blurred \\ vision\end{tabular}          &-                  &-                  &-                 &-                 & \multicolumn{1}{l|}{\cellcolor[HTML]{C0C0C0}-} &-                                              &-                                              &- & \multicolumn{1}{l|}{\cellcolor[HTML]{9B9B9B}-} &-                                              &-                                              &- \\ \hline
\begin{tabular}[c]{@{}l@{}}12. Dizzy \\ (eyes open)\end{tabular}       &-                  &-                  &-                 &-                 &-                                              &-                                              &-                                              &- & \multicolumn{1}{l|}{\cellcolor[HTML]{9B9B9B}-} &-                                              &-                                              &- \\ \hline
\begin{tabular}[c]{@{}l@{}}13. Dizzy \\ (eyes closed)\end{tabular}     &-                  &-                  &-                 &-                 &-                                              &-                                              &-                                              &- &-                                              &-                                              &-                                              &- \\ \hline
14. Vertigo                                                            &-                  &-                  &-                 &-                 &-                                              &-                                              &-                                              &- &-                                              &-                                              &-                                              &- \\ \hline
\begin{tabular}[c]{@{}l@{}}15. Stomach \\ awareness\end{tabular}       &-                  &-                  &-                 &-                 &-                                              & \multicolumn{1}{c|}{\cellcolor[HTML]{C0C0C0}-} &-                                              &- &-                                              & \multicolumn{1}{c|}{\cellcolor[HTML]{9B9B9B}-} & -                                              &- \\ \hline
16. Burping                                                            &-                  &-                  &-                 & -                 &-                                              &-                                              &-                                              &- &-                                              &-                                              &-                                              &- \\ \hline
\end{tabular}
\end{table}


\begin{table}[ht]
\centering
\caption{Weights for Symptoms}
\label{table3}
\begin{tabular}{|l|l|l|l|}
\hline
\multicolumn{4}{|c|}{Weights}                                                                                                 \\ \hline
\textit{Symptoms}           & \cellcolor[HTML]{EFEFEF}Nausea & \cellcolor[HTML]{EFEFEF}Oculomotor & \cellcolor[HTML]{EFEFEF}Disorientation \\ \hline
1. General discomfort       & \cellcolor[HTML]{EFEFEF}1      & \cellcolor[HTML]{EFEFEF}1          & \cellcolor[HTML]{EFEFEF}               \\ \hline
2. Fatigue                  & \cellcolor[HTML]{EFEFEF}       & \cellcolor[HTML]{EFEFEF}1          & \cellcolor[HTML]{EFEFEF}               \\ \hline
3. Headache                 & \cellcolor[HTML]{EFEFEF}       & \cellcolor[HTML]{EFEFEF}1          & \cellcolor[HTML]{EFEFEF}               \\ \hline
4. Eye strain               & \cellcolor[HTML]{EFEFEF}       & \cellcolor[HTML]{EFEFEF}1          & \cellcolor[HTML]{EFEFEF}               \\ \hline
5. Difficult focusing       & \cellcolor[HTML]{EFEFEF}       & \cellcolor[HTML]{EFEFEF}1          & \cellcolor[HTML]{EFEFEF}1              \\ \hline
6. Increased salivation     & \cellcolor[HTML]{EFEFEF}1      & \cellcolor[HTML]{EFEFEF}           & \cellcolor[HTML]{EFEFEF}               \\ \hline
7. Sweating                 & \cellcolor[HTML]{EFEFEF}1      & \cellcolor[HTML]{EFEFEF}           & \cellcolor[HTML]{EFEFEF}               \\ \hline
8. Nausea                   & \cellcolor[HTML]{EFEFEF}1      & \cellcolor[HTML]{EFEFEF}           & \cellcolor[HTML]{EFEFEF}1              \\ \hline
9. Difficulty concentrating & \cellcolor[HTML]{EFEFEF}1      & \cellcolor[HTML]{EFEFEF}1          & \cellcolor[HTML]{EFEFEF}               \\ \hline
10. "Fullness of head"      & \cellcolor[HTML]{EFEFEF}       & \cellcolor[HTML]{EFEFEF}           & \cellcolor[HTML]{EFEFEF}1              \\ \hline
11. Blurred vision          & \cellcolor[HTML]{EFEFEF}       & \cellcolor[HTML]{EFEFEF}1          & \cellcolor[HTML]{EFEFEF}1              \\ \hline
12. Dizzy (eyes open)       & \cellcolor[HTML]{EFEFEF}       & \cellcolor[HTML]{EFEFEF}           & \cellcolor[HTML]{EFEFEF}1              \\ \hline
13. Dizzy (eyes closed)     & \cellcolor[HTML]{EFEFEF}       & \cellcolor[HTML]{EFEFEF}           & \cellcolor[HTML]{EFEFEF}1              \\ \hline
14. Vertigo                 & \cellcolor[HTML]{EFEFEF}       & \cellcolor[HTML]{EFEFEF}           & \cellcolor[HTML]{EFEFEF}1              \\ \hline
15. Stomach awareness       & \cellcolor[HTML]{EFEFEF}1      & \cellcolor[HTML]{EFEFEF}           & \cellcolor[HTML]{EFEFEF}               \\ \hline
16. Burping                 & \cellcolor[HTML]{EFEFEF}1      & \cellcolor[HTML]{EFEFEF}           & \cellcolor[HTML]{EFEFEF}               \\ \hline
\end{tabular}
\end{table}

\section{Conclusion}
\label{sec:7}

This work is motivated by the widespread use of immersion devices such as HMD in virtual environments and the need of strategies to reduce the level of visual discomfort caused by such devices. 
Among several discomfort-causing factors, our study focused on suggesting design guidelines with an special attention to simulation of human vision focus of attention.
More specifically,  our proposal design a guideline that helps developers resolves the basics issues that causes discomfort in HMDs.
The dynamic focus model mentioned in this work was used to develop a first-person space shooter integrated with the dynamic focus selection heuristic, earlier created by us. 
We have developed the first automatic selection of focus game for HMD without eye tracking systems. The mechanic and gameplay is made in order to extrapolate the focus selection importance.
Besides that, our proposed test protocol can contribute to researches to evaluate their project’s users or companies seeking solutions to reduce discomfort in these devices.
We believe that there is still space for the development of new  games that uses this strategy in order to create a more comfortable experience to users and improve contribution guideline. In that sense, a public domain project was development for the Unity3D game engine driven by the proposed concepts. This project allows developers to create variations projects and adjust their own heuristic as needed.

\nolinenumbers

\bibliography{library}

\bibliographystyle{abbrv}

\end{document}